# Unsupervised CP-UNet Framework for Denoising DAS Data with Decay Noise

Tianye Huang,[1,2,*] Aopeng Li,[1] Xiang Li,[1] Jing Zhang,[1] Sijing Xian,[5] Qi Zhang,[7] Xiangyun Hu,[1] Mingkong Lu,[3] Liangming Xiong,[6] and Guodong Chen,[4]

[1] School of Mechanical Engineering and Electronic Information, China University of Geosciences (Wuhan), Wuhan, Hubei, China
[2] Shenzhen Research Institute of China University of Geosciences, Shenzhen, Guangdong, China
[3] Yazheng Technology Group Co., Ltd, Wuhan, Hubei, China
[4] China Electric Power Planning & Engineering Institute, Beijing, China.
[5] Electronics, Electrical Appliances, and Intelligent Connectivity Department, Liuzhou Wuling New Energy Automobile Co., Ltd
[6] Yangtze optical fiber and cable Joint Stock Limited Company, Wuhan, Hubei, China
[7] School of Economics and Management, China University of Geosciences (Wuhan), Wuhan, Hubei, China

*Corresponding author: Tianye Huang (E-mail:*huangty@cug.edu.cn)

## Abstract

Distributed acoustic sensor (DAS) technology leverages optical fiber cables to detect acoustic signals, providing cost-effective and dense monitoring capabilities. It offers several advantages including resistance to extreme conditions, immunity to electromagnetic interference, and accurate detection. However, DAS typically exhibits a lower signal-to-noise ratio (S/N) compared to geophones and is susceptible to various noise types, such as random noise, erratic noise, level noise, and long-period noise. This reduced S/N can negatively impact data analyses containing inversion and interpretation. While artificial intelligence has demonstrated excellent denoising capabilities, most existing methods rely on supervised learning with labeled data, which imposes stringent requirements on the quality of the labels. To address this issue, we develop a label-free unsupervised learning (UL) network model based on Context-Pyramid-UNet (CP-UNet) to suppress erratic and random noises in DAS data. The CP-UNet utilizes the Context

Pyramid Module in the encoding and decoding process to extract features and reconstruct the DAS data. To enhance the connectivity between shallow and deep features, we add a Connected Module (CM) to both encoding and decoding section. Layer Normalization (LN) is utilized to replace the commonly employed Batch Normalization (BN), accelerating the convergence of the model and preventing gradient explosion during training. Huber-loss is adopted as our loss function whose parameters are experimentally determined. We apply the network to both the 2-D synthetic and filed data. Comparing to traditional denoising methods and the latest UL framework, our proposed method demonstrates superior noise reduction performance.

## 1. Introduction

Distributed acoustic sensing (DAS) is a cutting-edge technology with the potential to transform seismic exploration [1]. DAS technology makes use of the sensitivity of backward Rayleigh scattered light in the optical fiber to changes in the physical quantity of the fiber. DAS utilizes the optical fiber as a sensor to sense the perturbation caused by seismic waves transmits the information of the perturbation to the signal acquisition terminal, and then recovers the perturbed signals through the demodulation of the backward Rayleigh scattered light signals. DAS technology offers distinct benefits over conventional seismic systems, effectively tackling key challenges in three critical areas [2]: (1) DAS transforms a single fiber into numerous virtual sensors, enabling continuous, real-time monitoring along its entire length. It removes the need for multiple physical sensors, streamlining the monitoring process. (2) The DAS system provides immediate data capture and pinpoints the exact positions of sensor activity along the fiber. (3) The DAS system delivers high sensitivity using common single-mode fiber, all at a low cost. Even though DAS has many benefits, its signal clarity is often reduced by different kinds of noise coming from both the surroundings and the equipment itself, such as horizontal noise [3], random noise [4], erratic noise [5], fading noise, and ringing noise. The presence of these noises can distort the seismic data, affecting its interpretation. Although different methods like spectrum extraction, remix (SERM), and sub-band phase-shift transform have been developed to enhance DAS performance, the signal-to-noise ratio (S/N) still needs further improvement [6].

Researchers have developed and improved numerous techniques to mitigate noise in DAS data. From the viewpoint of time-frequency filtering, The high-frequency and unstable noise is attenuated using a combination of a band-pass filter and a median filter. [7]. A cascade framework was proposed to attenuate high-frequency, unstable, and horizontal noise in DAS data by combining different simultaneous filtering methods. [8]. By using multi-scale analysis based on wavelet transform [9], the method achieves the optimal sparse representation of DAS data through threshold switching. Then, appropriate threshold functions are set based on the differences in the properties of effective signals at different scales to determine the frequency domain coefficients of the DAS signal. Finally, the denoised DAS signal is obtained by performing an inverse transformation on the coefficients. Researchers proposed the curvelet transform and the shearlet transform to effectively attenuate the random noise [10]. The modal decomposition method can decompose the DAS signal into different modes, and get a useful signal by selecting the modes related to the valid signal. The empirical mode decomposition (EMD) [11] and the variational mode decomposition (VMD) [12] results in random and high-frequency noise attenuation. Traditional noise reduction methods work well for random or Gaussian noise but struggle with the complexities of DAS data.

Deep learning stands out in seismic studies for its adaptability and strong ability to analyze complex data. It automatically extracts key features from data using sophisticated neural networks, enabling precise pattern recognition and signal analysis. Deep learning is used for geophysics applications such as coherent noise attenuation [13], seismic inversion [14], and seismic interpolation. Deep learning also enables the precise detection and separation of noise in DAS data, clearly differentiating it from the actual signals. Researchers enhanced the deep neural network structure to represent the signal more accurately and efficiently for DAS data [15]. Researchers employ a supervised network centered on densely connected residuals for denoising complex noise [16]. However, the conventional supervised learning methods require many labels for training the network, which is not cost-effective and very time-consuming. While unsupervised learning offers a label-free solution for DAS data noise reduction. For example, Researchers designed deep-denoising unsupervised learning (DDUL) and a fully connected neural network for DAS noise suppression [17]. Researchers designed an attention cycle GAN network (ACGNet) based on unpaired datasets to train the network [18]. However, for the unsupervised learning network, the performance of noise reduction still needs improvement.

In this paper, we introduce a Context Pyramid-UNet (CP-UNet) network architecture, developed using unsupervised learning, for DAS data noise reduction. The encoding and decoding in the UNet architecture use the Context Pyramid Module to extract data features from shallow to deep layers comprehensively and complete DAS data reconstruction. Considering the physical properties of the noise, the loss function is chosen to be Huber-loss with l1 and l2 losses. By testing the CP-UNet in both synthetic and real-time datasets, it shows great potential for attenuating complex noise. By comparing to two basic methods and a state-of-the-art deep learning network, it is verified that the proposed CP-UNet method is effective and superior.

## 2. Method

The CP-UNet utilizes four layered encoders to capture features ranging from superficial to profound within the DAS data. In the decoding process, four stacked decoders are used to reconstruct the useful signal. It is assumed that Y is the data with noise and X is the useful signal hidden in the noise. This relationship is represented by the equation below.

$$X = Y - N \tag{1}$$

where N represents complex noise in the DAS data, and Y represents the raw noisy DAS data. CP-UNet is designed with a Context Pyramid Module (CPM) and a Connectivity Module (CM), aiming to extract pure hidden signals from cluttered backgrounds. The network architecture shown in Fig. 1 mainly consists of CPM and CM. CPM and CM mainly consist of linear layer normalization (LN), leaky rectified linear unit (LeakyReLU), and Dropout. When the input one-dimensional data passes through the linear in the lth layer of CPM and CM, the output is represented as

$$F_l = A_l Y_l + b_l \tag{2}$$

where $A_l$ represents weight matrices and $b_l$ represents bias vectors of the $l$th CPM, respectively, and $Y_l$ represents $l$th input. Layer normalization (LN) is used to accelerate the convergence speed and generalization ability of the model while preventing gradient explosion. The addition of an LN layer after the Linear layer effectively maintains specific relationships between different features and improves the effectiveness of the training process. LN normalizes the input data to obtain a stable distribution, which reduces the risk of overfitting, makestraining more stable, and accelerates network convergence. After the output of the linear layer, the LN output of the $l$th CPM is given as

$$LN(F_l) = \gamma \frac{F_l - \mu(F_l)}{\sqrt{\sigma^2(F_l) + \varepsilon}} + \beta \tag{3}$$

where $\mu(F_l)$ and $\sigma^2(F_l)$ denote the mean and variance of the $l$th layer, $\beta$ and $\gamma$ denote $l$th learning parameters. $\varepsilon$ denotes a small value to ensure numerical stability. To improve the nonlinear mapping learning ability of the neural network, LeakReLU is used as the activation function of the network. LeakyReLU transforms the output of the LN in a nonlinear way, and the output is

$$r_l = \max(\lambda LN(F_l), LN(F_l)) \tag{4}$$

where $\lambda$ denotes a fixed value (0.2). As the CP-UNet is a complex deep learning model, it can sometimes overfit during training. To address this issue, we incorporated a dropout layer following by the LeakReLU layer, enhancing the model's ability to reduce noise effectively.

The CPM is shown in Fig. 2. Specifically, CPM performs linear transformations of different dimensions on the input feature dimensions. Next, for features of different dimensions, a linear transformation, Layer Normalization (LN), Leaky ReLU, and dropout are applied. Then, the features from three different dimensions are fused through concatenation, integrating multi-scale features. This process allows the network to capture signal features from different dimensions and extract signals hidden in noise. The CPM result after the $l$th layer is

$$C_l = Concat(F_{l1}, o_{l1}, o_{l2}) \tag{5}$$

where $F_{l1}$ denotes the branch with unchanged feature dimensions after linear. The $o_{l1}$ and $o_{l2}$ denote the branch where the feature dimensions change after linear. To enhance the feature transfer capability of the network, we incorporated a Connection Module (CM) that links the corresponding stages of the encoder and decoder within the network. The CM consists of linear, LN, LeakyReLU, and dropout, which is shown in Fig. 2. By directly connecting shallow and deep features, the CM mitigates the vanishing gradient problem while integrating multi-level features. This allows the decoder to leverage both shallow and deep features, thereby improving the denoising performance of the model. The CM result after the $l$th layer is LeakyReLU(LN($F_l$)). The CP-UNet model serves as a denoiser, establishing a complex relationship between the noisy input data $y$ and the denoised data $x$. Denoised data can be expressed as

$$x = P(y; \Theta) \tag{6}$$

where P(·) denotes CP-Unet. $\Theta$ denotes the parameters in the network. In this study, considering the physical characteristics of the noise, The Huber loss is utilized as our loss function, integrating the benefits of both mean square error (MSE) and mean absolute error (MAE), thereby offering increased robustness to outliers. The formula for this function is as follows

$$Loss_{huber} = \begin{cases} \frac{1}{2}(x-y)^2, & for |x-y| < \alpha \\ \alpha(x-y) - \frac{1}{2}\alpha^2, & otherwise \end{cases} \tag{7}$$

where $a$ denotes the penalty for outliers in an asymptotic manner, which will be discussed subsequently. In order to obtain the most performance of $\Theta$, we use the Adam optimizer to update the network's parameters.

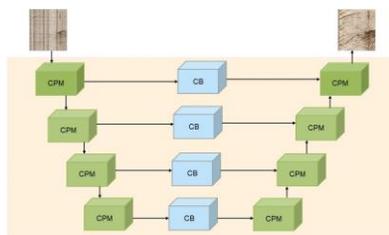

Fig. 1. Architectural diagram of CP-UNet.

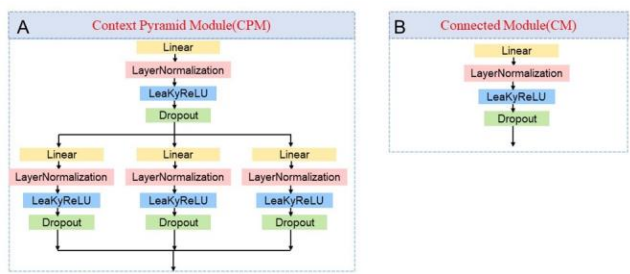

Fig. 2. Architectural details diagram of CP-UNet, where boxes A and B denote the CPM and CM, respectively.

The parameters of the loss function determine the performance of the CP-UNet denoiser. For Huber loss, the threshold α has a critical impact on the loss function, which determines how errors are handled in different ranges. When the error is less than the threshold, Huber loss behaves as Mean Squared Error (MSE), applying a quadratic penalty to small errors, which aids in optimization stability and rapid convergence. When the error exceeds the threshold, the loss function transitions to Mean Absolute Error (MAE), applying a linear penalty to large errors, thereby mitigating the influence of outliers on the overall loss and enhancing the model's robustness. In this study, we set the threshold $a$ to five different values and obtained five different. The S/N is tested on the same dataset with the parameter $a$ set to 1.0, 1.1, 1.2, 1.3, and 1.4, respectively. The corresponding S/N results were 10.47 dB, 10.97 dB, 11.25 dB, 11.02 dB, and 10.79 dB. The results indicate that the denoiser performs best when the threshold of 1.2.

The CP-UNet framework, an adaptation of the UNet design, balances processing speed with the ability to effectively extract data features. This method segments noisy DAS data into smaller pieces for preprocessing [19]. The segments are labeled using set parameters for size and overlap. This approach ensures efficient data management while preserving the quality of feature extraction. The DAS data is processed sequentially, moving in a grid pattern according to set patch dimensions. To ensure a continuous waveform as the analysis window moves, we set the patch size and slide size according to specific calculations involving constants C x (C-D), respectively. After patching, the data is transformed into one-dimensional and then input into the network for noise reduction.

In the synthetic data experiments, the output feature dimensions of the 4 CPM and CM in the encoder are 64, 32, 16, and 8. The output feature dimensions of the 4 CPM in the decoder are 8, 16, 32, and 64. In the filed data experiments, the output feature dimensions of the 4 CPM in the encoder and CM output feature dimensions are 128, 64, 32, and 16.

The 4 CPM output feature dimensions in the decoder are 16, 32, 64, 128. The training process is presented as follows:

Step 1: Choose 48x48 signal patches and input them into the network.

Step 2: Compute the loss, apply gradient descent to adjust it, and use the Adam optimizer to refine the model's parameters.

Step 3: Repeat steps 1 and 2 for a total of 100 training epochs.

Step 4: Use the trained model to predict clean signals from the noisy seismic data.

## 3. Synthetic experiments

The performance of deep learning frameworks relies on accurate, high-quality training sets. Noise-free data is difficult to obtain in practical exploration and we use forward modeling to obtain noise-free DAS data. To generate the model, we performed forward velocity modeling with different layers (flat, inclined, concave, etc.). Fig. 3 presents a 2D geological model, which spans 256 units horizontally and extends to a depth of 512 units. Inverted triangles within the diagram indicate the locations of seismic sources. The spatial sampling interval is set to 1 m and the sampling frequency is 1000 Hz. Seismic waves arise from sharp ruptures and activity in the earth's internal material. Seismic waves propagate in a solid medium and show obvious elastic properties. Therefore, the model uses elastic waves to simulate these signals in this study with the specific parameters shown in Table I.

The noise dataset consists of synthetic noise and field noise, with synthetic noise accounting for 85% and real noise accounting for 15%. Synthetic noise primarily comprises random noise. A cascade framework was proposed to encourage noise attenuation in the FORGE dataset [15]. We use this cascade framework to extract random and erratic noise and then add it and synthetic noise to the noise-free data to form noisy data. The S/N (signal-to-noise ratio) of the noise-containing data ranges from -0.5 dB to 1.5 dB.

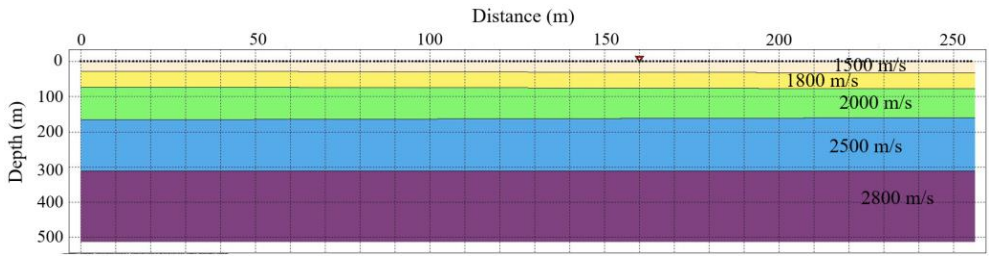

Fig. 3. 2-D geological model for forward modeling.

TABLE I

PARAMETERS OF FORWARD MODELING

| Seismic wavelet | Single, ricker |
|---|---|
| The dominant frequency of the seismic wavelet | 30-75 Hz |
| Trace interval | 1 m |
| Wave equation | Elastic |

| Sampling interval | 0.001 s |
| --- | --- |
| Wave velocity (m/s) | 1500-4000 m/s |
| Density | 1900-2300 kg/m³ |

The S/N was used to measure the success of noise reduction, as expressed in Eq.8.

$$S/N = 20\log_{10} \frac{\| X \|}{\| X - \hat{X} \|_2^2} \tag{8}$$

where $X$ and $\hat{X}$ represent the clean signal and the noisy signal, respectively. The clean signal is the term we use for the noise-free version of the data in our study. We measure the model's noise reduction effectiveness by comparing the S/N before and after processing.

In our synthetic data, we've added various types of noise, such as random and erratic sounds, as shown in Fig. 4c. Random noise includes sounds like wind, rain, and traffic, and is known for being unpredictable and incoherent. This can create occasional unusual data points. Erratic noise, primarily stemming from environmental and geological events, has a broad frequency range. Our synthetic data includes two types of noise: erratic and Gaussian random noise. Gaussian noise is created by software, whereas erratic noise comes from real DAS signal recordings, as shown in Fig. 4a. To tackle noise, we apply four different noise reduction methods to the affected DAS data and evaluate their performance using the S/N. Fig. 5b shows the results of the band-pass filter and median filter. Fig. 5c and 5d show the noise reduction results for DDUL and CP-UNet methods. The analysis reveals that band-pass filtering still leaves behind noticeable erratic and random noise, and the median filter also retains some random noise. In comparison, CP-UNet proves to be much more successful at removing noise. However, from Fig. 6a-d, a portion of the DAS signal was selected with a rectangular box and amplified. Fig. 6e-h show the amplified portion of the signal. In Fig. 6g, it is observed that there is signal leakage from DDUL, while the CP-UNet retains the signal component completely as shown in Fig. 6h. The S/Ns obtained by the four methods are 6.48 dB, 7.05 dB, 8.75 dB, and 11.25 dB, respectively.

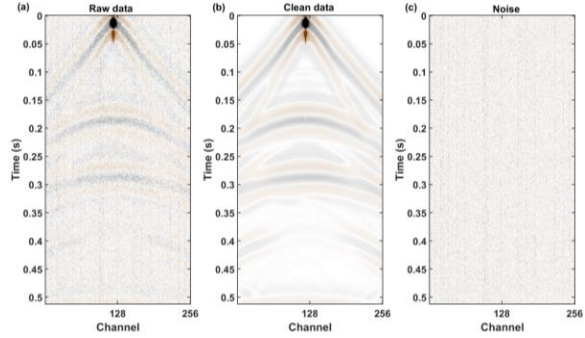

Fig. 4. The 2D DAS synthetic data incorporates two varieties of noise: random noise, and erratic noise. (a) noise data, (b) clean data, (c) noisy part.

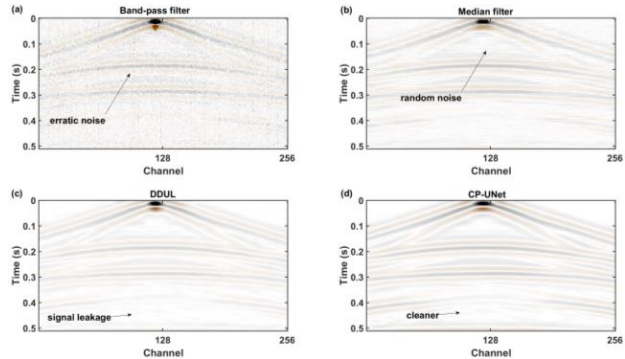

Fig. 5. Denoising results on a synthetic dataset using four distinct Methods. (a) Band-pass filter, (b) Median filter, (c) DDUL, (d) CP-UNet.

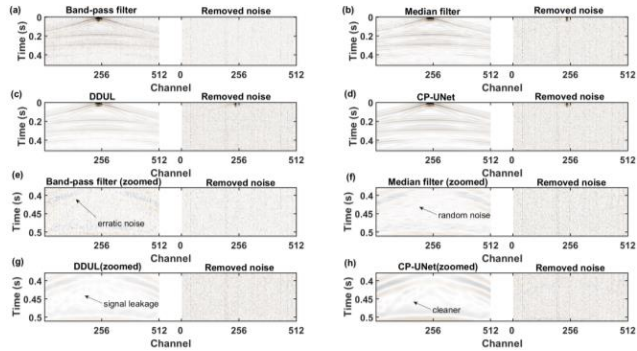

Fig. 6. Denoised data using (a) Band-pass filter, (b) Median filter, (c) DDUL, and (d) CP-UNet. (e)-(h) are the enlarged rectangular boxes of a-d, respectively.

## 4. Filed experiments

In this section, two FORGE seismic datasets are selected to test the network performance of CP-UNet. The program collected seismic records for about 10 days in phase 2C. The sample size of the test data was 2000x960. each test data was Patch into 48x48 size. As the synthetic data, we chose the bandpass filter, median filter, and DDUL model for comparison. The DDUL model is also patched and has the same patch size and slide size. For the FORGE dataset, Fig. 7 shows the denoising results of different methods.

From Fig. 7a, the band-pass filter still has erratic noise, and Fig. 7b shows that the median filter also has erratic noise. Although DDUL can eliminate random noise, erratic noise still exists. Compared with the three benchmark models, CP-UNet attenuates both random noise and erratic noise, leaving useful signals. Fig. 8 shows zooming in on a portion of the data to observe the signal in detail, it can be seen that CP-UNet can successfully extract useful signals from highly contaminated noisy data.

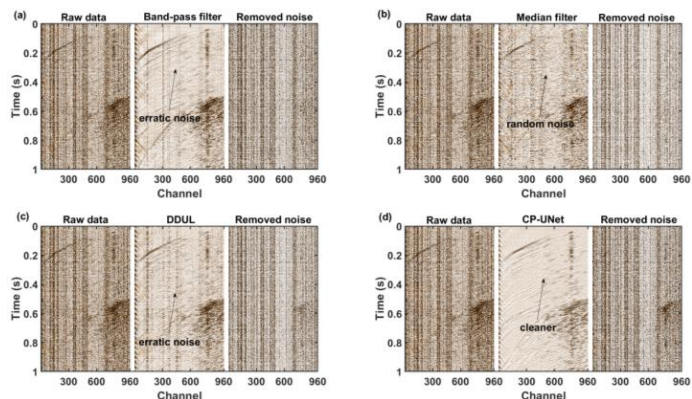

Fig. 7. Comparative analysis of denoising techniques on test set 1 for (a) Band-pass filter, (b) Median filter, (c) DDUL, and (d) CP-UNet.

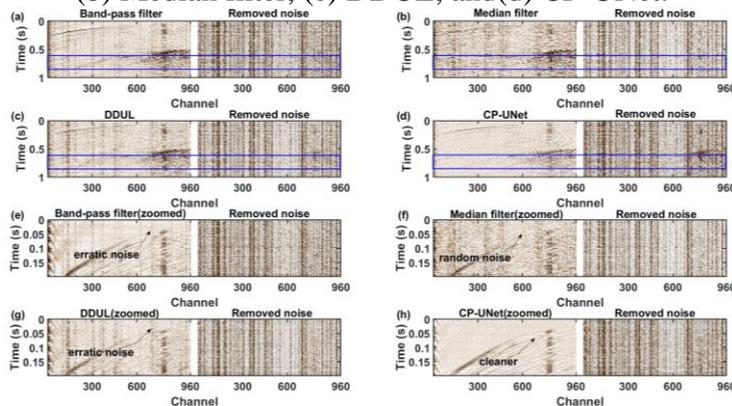

Fig. 8. Comparative analysis of denoising techniques on test set 1 for (a) Band-pass filter, (b) Median filter, (c) DDUL, (d) CP-UNet, (e) expanded cross-sectional view of a blue box in Fig. 7a, (f) expanded cross-sectional view of a blue box in Fig. 7b, (g) expanded cross-sectional view of a blue box in Fig. 7c and (h) expanded cross-sectional view of a blue box in Fig. 7d.

## 5. Conclusion

This paper proposes an unsupervised CP-UNet denoiser that can improve the signal-to-noise ratio in the absence of a large number of labels. The CP-UNet consists mainly of a CPM and a CM. The CPM extracts features and reconstructs DAS data during encoding and decoding, while the CM enhances the connectivity between shallow and deep features.

Performance was evaluated through denoising experiments on both synthetic and real-time field data, comparing outcomes across various methodologies. For data preprocessing, we applied patching and un-patching techniques to generate a diverse array of small patches from the noisy input data. Filed data testing results indicate that the network architecture effectively and significantly reduces noise in data characterized by random and erratic noise types. Furthermore, the proposed network outperforms traditional methods in terms of S/N metrics. Additionally, our framework is adept at detecting hidden signals amid highly noisy environments. Our findings underscore the strong capability of this approach to enhance the clarity of seismic data, which in turn boosts the accuracy of seismic forecasts.

**Funding.** National Natural Science Foundation of China (42327803); Guangxi Science and Technology Major Program(AA23062024); Open Fund of SINOPEC Key Laboratory of Geophysics; Open Project Program of Wuhan National Laboratory for Optoelectronics (2023WNLOKF007); Major Science and Technology Projects of Wuhan (2023010302020030); Guangdong Basic and Applied Basic Research Foundation (2023A1515010965, 2024A1515010017).

# References

1. J. Johny, S. Amos, and R. Prabhu, "Optical Fibre-Based Sensors for Oil and Gas Applications," *Sensors.*, vol. 21, no. 18, pp. 66047, 2021.
2. G. Binder, A. Titov, Y. Liu, J. Simmons, A. Tura, G. Byerley, and D. Monk, "Modeling the seismic response of individual hydraulic fracturing stages observed in a time-lapse distributed acoustic sensing vertical seismic profiling survey," *Geophysics.*, vol. 85, no. 4, pp.T225-T235, 2020.
3. T. Zhong, M. Cheng, S. Lu, X. Dong, and Y. Li, "RCEN: a deep-learning-based background noise suppression method for DAS-VSP records," *IEEE Geoscience and Remote Sens9ing Letters.*, vol. 19, pp. 1-5, 2021.
4. H. Wang, Q. Zhang, G. Zhang, J. Fang, and Y. Chen, "Self-training and learning the waveform features of microseismic data using an adaptive dictionary," *Geophysics.*, vol. 85, no. 3, pp. KS51-KS61, 2020.
5. W. Jeong, M. S. Almubarak, and C. Tsingas, "Seismic erratic noise attenuation using unsupervised anomaly detection," *Geophysical Prospecting*, vol. 69, no. 7, pp. 1473-1486, 2021.
6. Cheng, Y., H. Wang, D. Li, Y. Qiu, M. Luo, X. Zhang, J. Zhang, Z. Wu, T. Huang, and X. Li, "Interference fading mitigation in coherent Φ-OTDR based on subband phase-shift transform," *IEEE Photonics Journal.*, vol. 15, no. 5, pp. 6802006, 2023.
7. A. Lellouch, R. Schultz, J. Lindsey, B. Biondi, and W. L. Ellsworth, "Low-magnitude seismicity with a downhole distributed acoustic sensing array — examples from the forge geothermal experiment," *Journal of Geophysical Research Solid Earth.*, vol. 126, no. 1, pp. e2020JB020462, 2021.
8. Y. Chen, A. Savvaidis, S. Fomel, Y. Chen, O. M. Saad, H. Wang, and W. Chen, "Denoising of distributed acoustic sensing seismic data using an integrated framework" *Seismological Society of America.*, vol. 94, no. 1, pp. 457-472, 2023.
9. D. Li, H. Wang, X. Wang, X. Li, T. Huang, M. Ge, J. Yin, S. Chen, B. Huang, K. Guan, C. He, H. Hu, K. Li, and Z. Lian, "Denoising algorithm of Φ-OTDR signal based on curvelet transform with adaptive threshold," *Optics Communications.*, vol. 545, pp. 129708, 2023.
10. X. Liang, Y. Li, and C. Zhang, "Noise suppression for microseismic data by non-subsampled shearlet transform based on singular value decomposition," *Geophysical Prospecting*, vol. 66, no .5, pp. 894-903, 2018.
11. J. Li, Y. Li, Y. Li, and Z. Qian, "Downhole microseismic signal denoising via empirical wavelet transform and adaptive thresholding," *Journal of Geophysics and Engineering*, vol. 115, no. 66, pp. 2469-2480, 2018.
12. Y. Zhang, Z. Haoran, Y. Yang, L. Naihao, and G. Jinghuai, "Seismic random noise separation and attenuation based on MVMD and MSSA," *IEEE Transactions on Geoscience and Remote Sensing.*, vol. 60, no. 4, pp. 1-16, 2021.
13. N. Jia, H. Ma, X. Dong, and Y. Li, "Background noise suppression using trainable nonlinear reaction diffusion assisted by robust principal component analysis," *Exploration Geophysics*, vol. 51, no. 6, pp. 642-651, 2020.
14. T. A. Larsen Greiner, J. E. Lie, O. Kolbjørnsen, A. Kjelsrud Evensen, E. Harris Nilsen, H. Zhao, V. Demyanov, and L.-J. Gelius, "Unsupervised deep learning with higher-order total-variation regularization for multidimensional seismic data reconstruction," *Geophysics.*, vol. 87, no. 2, pp. V59-V73, 2022.
15. L. Yang, S. Fomel, S. Wang, X. Chen, and Y. Chen, "Denoising distributed acoustic sensing data using unsupervised deep learning," *Geophysics.*, vol. 88, no. 4, pp. V317-V332, 2023.
16. T. Huang, A. Li, D. Li, J. Zhang, X. Li, L. Xiong, J. Tu, W. Sun, and X. Hu, "Multiple noise reduction for distributed acoustic sensing data processing through densely connected residual convolutional networks," *Journal of Applied Geophysics.*, vol. 228, pp. 105464, 2024.
17. L. Yang, S. Wang, S. Wang, O. M. Saad, O. M. Saad, Y. A. S. I. Oboue, and Y. Chen, "Unsupervised 3-D random noise attenuation using deep skip autoencoder," *IEEE Transactions on Geoscience and Remote Sensing.*, vol. 60, pp. 1-16, 2021.
18. H. Ma, H. Ba, Y. Li, Y. Zhao, and N. Wu, "Unpaired training: Optimize the seismic data denoising model without paired training data," *Geophysics.*, vol. 88, no. 1, pp. WA345-WA360, 2023.
19. Y. Chen, M. Zhang, M. Bai, and W. Chen, "Improving the signal-to-noise ratio of seismological datasets by unsupervised machine learning," *Geophysics.*, vol. 90, no. 4, pp. 1552-1564, 2019.